\documentclass[english]{article}
\usepackage[T1]{fontenc}
\usepackage[latin9]{inputenc}
\usepackage{babel}
\usepackage{amsmath}
\usepackage{stackrel}
\usepackage{graphicx}
\usepackage{geometry}
\usepackage[]
 {hyperref}

\makeatletter
\numberwithin{equation}{section}

\makeatother

\begin{document}
\title{Confidence Intervals for the Savitzky-Golay Filter with\\
an Application to the Keeling Data for Atmospheric CO\textsubscript{2}}
\author{Paul W. Oxby}
\maketitle
\begin{abstract}
The Savitzky-Golay FIR digital filter is based on a least-squares
polynomial fit to a sample of equally spaced data. The polynomial
fit gives the filter the ability to preserve moments of features in
the data like peak width. However the S-G filter is not generally
regarded as having a sound statistical basis. This puts the filter
in the category of smoothing filters where the degree of smoothing
depends on the somewhat arbitrary choice of the filter parameters.
This arbitrariness makes the variance of the residuals between the
filter input and output an unreliable estimate of the variance of
the noise in the filter input. And without a reliable estimate of
the input noise variance there is no basis for determining statistically
meaningful confidence intervals on the filter output.

This paper proposes a method of using the S-G filter to determine
a reliable estimate of the variance of the noise in the data. This
estimate is then used as the basis for selecting appropriate filter
parameters and determining statistically meaningful confidence intervals
on the filter output. To illustrate the proposed method an analysis
of the Keeling measurements of atmospheric CO\textsubscript{2} concentration
is presented.
\end{abstract}

\section{Introduction}

In a 1964 paper Savitzky and Golay {[}1{]} showed how a least-squares
polynomial fit to a sample of equally spaced data could be used as
the basis of an FIR digital filter to smooth noisy data. This filter
is popular because it is better at preserving data features like peak
width than a simple moving average. Despite the popularity of the
Savitzky-Golay (S-G) filter it is not generally regarded as a rigorous
statistical method. For example, Press, et al. {[}2{]} write of the
S-G filter: ``the smoothing of data lies in a murky area, beyond
the fringe of some better posed, and therefore more highly recommended
{[}statistical{]} techniques.'' One desirable characteristic of a
rigorous statistical technique is that it provides an estimate of
the uncertainty in the results. This estimate is often expressed in
terms of a confidence interval. In their paper Savitzky and Golay
do not consider confidence intervals and neither do some textbooks
{[}2{]} {[}3{]} which provide an otherwise detailed description of
the S-G filter.

In least-squares parameter estimation confidence intervals are based
on an estimate of the uncertainty in the data. Subject to standard
assumptions regarding the nature of the uncertainty, this estimate
of the uncertainty is based on the minimized sum of squares of residuals.
This poses a problem with a nonparametric smoothing filter. In the
case of the S-G filter both the size of the sample interval and the
degree of the fitting polynomial can be adjusted to give various degrees
of smoothing. Various degrees of smoothing give inconsistent estimates
of the uncertainty in the data. And these inconsistent estimates are
unsuitable for determining confidence intervals. This is what places
the S-G filter ``in a murky area, beyond the fringe of some better
posed ... techniques.''

This paper describes a heuristic method for using the S-G filter to
produce a reasonably consistent estimate of the uncertainty in the
data. This estimate then serves as a guide to selecting appropriate
filter parameters. And with appropriate S-G filter parameters reliable
confidence intervals on the filter output can be determined thereby
rescuing the S-G filter from ``a murky area, beyond the fringe.''
To illustrate this method it is applied to an analysis of the Keeling
data for atmospheric CO\textsubscript{2} concentration measured over
a 67 year period {[}4{]}.

\section{The Savitzky-Golay Smoothing Filter}

This section shows how the Savitzky-Golay (S-G) filter is derived
from the statistical theory of the optimal polynomial fit to a sample
of measurements corrupted by random noise. The signal processing terms
'signal' and 'noise' are used here because they are more general than
the customary statistical terms 'true value' and 'measurement error'.
The values of a hypothetical dependent variable, \emph{y}, are associated
with \emph{$2m+1$} equally spaced values of an independent variable,
\emph{x}. Without loss of generality the values of \emph{x} are scaled
to the interval $\left[-1,1\right]$. Therefore the spacing between
the values of \emph{x} is $1/m$:
\begin{equation}
x_{i}=\frac{i-1-m}{m}\quad(\mathrm{for}\:i=1\,...\,2m+1)
\end{equation}
It is further assumed that the signal can be well represented by a
polynomial with \emph{n} terms (i.e., degree $n-1$) over the range
of the \emph{$2m+1$} samples. To illustrate, if a quadratic equation
in \emph{x} with three unknown coefficients, $a_{1}$, $a_{2}$, and
$a_{3}$ is to be fit to five corresponding measurements of \emph{y}
then this can be expressed by the following set of five equations
in three unknowns: 
\begin{align}
y_{1} & =a_{1}+a_{2}\,x_{1}+a_{3}\,x_{1}^{2}\nonumber \\
y_{2} & =a_{1}+a_{2}\,x_{2}+a_{3}\,x_{2}^{2}\nonumber \\
y_{3} & =a_{1}+a_{2}\,x_{3}+a_{3}\,x_{3}^{2}\\
y_{4} & =a_{1}+a_{2}\,x_{4}+a_{3}\,x_{4}^{2}\nonumber \\
y_{5} & =a_{1}+a_{2}\,x_{5}+a_{3}\,x_{5}^{2}\nonumber 
\end{align}
So in this case $m=2$ and $n=3$. Equations 2.2 are more compactly
expressed in matrix notation as:

\begin{equation}
y=Xa
\end{equation}
where the elements of the matrix \emph{X} are given by:

\begin{equation}
X_{i,j}=x_{i}^{j-1}\quad(\mathrm{for}\:i=1\,...\,2m+1,\:j=1\,...\,n)
\end{equation}

Because there are more equations for the elements of vector \emph{a}
than there are elements of \emph{a} there won't be a solution for
\emph{a} that exactly satisfies all of the equations if there is random
noise associated with the dependent variable, \emph{y}. The system
of equations is said to be over-determined. For a given set of values
of the vector \emph{a} the residual vector, \emph{r}, is defined as:

\begin{equation}
r=y-Xa
\end{equation}

This optimal values of the elements of vector \emph{a} are obtained
by minimizing the weighted sum of squares of the residuals, $r^{T}Wr$,
with respect to the elements of \emph{a} where the matrix \emph{W}
is a $2m+1$ by $2m+1$ diagonal matrix of weights. Premultiplying
both sides of Equation 2.3 by $X^{T}W$ gives a set of linear equations
that are not over-determined and whose solution for \emph{a} minimizes
the weighted sum of squares of the residuals, $r^{T}Wr$:

\begin{equation}
a=\left(X^{T}WX\right)^{-1}X^{T}Wy
\end{equation}
The values of the polynomial fit are given by the vector $\widehat{y}$
whose elements are the best estimates of the values of the signal:

\begin{equation}
\widehat{y}=Xa=X\left(X^{T}WX\right)^{-1}X^{T}Wy
\end{equation}
If a particular element of the $\widehat{y}$ vector is of interest,
say the \emph{j}th element, then that element can be isolated from
$\widehat{y}$ by a simple matrix operation. The vector \emph{u} is
constructed with \emph{$2m+1$} elements all of which are set to zero
except for the \emph{j}th element which is set to one:

\begin{equation}
u_{i}=\left\{ \begin{array}{l}
0\;\mathrm{for}\:i\neq j\\
1\;\mathrm{for}\:i=j
\end{array}\right.\quad(\mathrm{for}\:i=1\,...\,2m+1)
\end{equation}

The \emph{j}th element of $\widehat{y}$ is now given by:

\begin{equation}
\widehat{y}_{j}=u^{T}\widehat{y}=u^{T}X\left(X^{T}WX\right)^{-1}X^{T}Wy
\end{equation}
The transpose of $u^{T}X\left(X^{T}WX\right)^{-1}X^{T}W$ is a vector
of \emph{$2m+1$} filter coefficients. These coefficients will be
denoted by $c(j)$ to emphasize that the values of the filter coefficients,
\emph{c,} depend on which of the $2m+1$ elements of the $\widehat{y}$
vector is of interest.

\begin{equation}
c(j)^{T}=u^{T}X\left(X^{T}WX\right)^{-1}X^{T}W\quad(\mathrm{for}\:u_{j}=1)
\end{equation}

Therefore the filtering operation is simply the dot product or convolution
of the filter coefficients, \emph{c}, with the measurements, \emph{y:}

\begin{equation}
\widehat{y}_{j}=c(j)^{T}y
\end{equation}

For the purpose of filtering, the optimal value of \emph{j} corresponds
to the middle element of $\widehat{y}$ with index $m+1$ and for
which $u_{m+1}=1$. When $j=m+1$ then \emph{$c(j)$} is a symmetric
vector and the filter is linear phase. However for the \emph{m} values
of \emph{y} at the beginning and end of the data set to be filtered
the sample interval cannot extend past the range of the data. For
these \emph{m} values the sample interval is fixed at the first and
last $2m+1$ values of \emph{y} and the value of \emph{j} ranges from
one to\emph{ m} for the first \emph{m} values of \emph{y} and from
$m+2$ to $2m+1$ for the last \emph{m} values of \emph{y}.

Equations 2.10 and 2.11 define the Savitzky-Golay smoothing filter.
The S-G filter can also be used to estimate the derivatives of the
signal. In particular, the estimate of the first derivative of the
signal is given by:

\begin{equation}
\frac{d\widehat{y}}{dx}=XDa=XD\left(X^{T}WX\right)^{-1}X^{T}Wy
\end{equation}
where \emph{D} is an \emph{n} by \emph{n} matrix whose $n-1$ nonzero
elements are based on the relation $dx^{k}/dx=k\,x^{k-1}$:

\begin{equation}
D_{k,k+1}=k/m\quad(\mathrm{for}\:k=1\,...\,n-1)
\end{equation}

Note that the independent variable, \emph{x}, is scaled to a spacing
of $1/m$ in Equation 2.1. It is convenient to make the derivative
independent of the sample size by dividing \emph{k} by \emph{m} in
this equation. So for the purpose of the derivative the effective
spacing of \emph{x} is one. From Equation 2.12 the vector of filter
coefficients, $cd$, giving the first derivative of $\widehat{y}_{j}$
is given by:

\begin{equation}
cd(j)^{T}=u^{T}XD\left(X^{T}WX\right)^{-1}X^{T}W\quad(\mathrm{for}\:u_{j}=1)
\end{equation}

\begin{equation}
\frac{d\widehat{y}_{j}}{dx}=cd(j)^{T}y
\end{equation}

Descriptions of the S-G filter often implicitly assign equal weights
to the residuals, \emph{r}, in minimizing the weighted sum of squares
of the residuals. In this case \emph{W} is an identity matrix. However
giving equal weights to the residuals does not optimize the smoothness
of the filter output. It has been shown {[}5{]} that, for a particular
measure of the smoothness of the filter output, the optimal weights
are given by a quadratic function of the weight indices or of \emph{x}
as defined by Equation 2.1:

\begin{equation}
W_{i,i}=\frac{3\,i}{2\,m+3}\left(2-\frac{i}{m+1}\right)=\frac{(m+1)^{2}-(m\,x_{i})^{2}}{(m+1)\,(2\,m+3)}\quad\left(\mathrm{for}\:i=1\,...\,2m+1\right)
\end{equation}

The mean of these weights is one and their value one sample interval
beyond the sample indices $\left[1,2m+1\right]$ is zero, i.e.:

\begin{equation}
W_{0,0}=W_{2m+2,2m+2}=0
\end{equation}

It is this particular smooth transition of the weights from zero to
their maximum value at the midpoint of the sample interval and back
to zero that optimizes the smoothness of the filter output.

\section{Confidence Intervals for the S-G Filter}

The output of a FIR filter, $\widehat{y}$, with $2m+1$ coefficients
is a convolution of the filter coefficients and the filter input (Equation
2.11):

\begin{equation}
\widehat{y}=c_{1}\,y_{1}+c_{2}\,y_{2}+\cdots+c_{2m+1}\,y_{2m+1}
\end{equation}

It is assumed that the noise in the filter input, \emph{e}, is an
independent and identically distributed (i.i.d.) random variable with
zero mean and variance $\sigma_{e}^{2}$. In least-squares parameter
estimation there is often a priori knowledge that a specific parametric
model accurately represents the signal in the data. In this case there
is little scope for over-fitting or under-fitting the data and the
following equation gives an estimate of the input noise variance:

\begin{equation}
\sigma_{e}^{2}=\frac{1}{q}\stackrel[i=1]{q}{\sum}\left(y_{i}-\widehat{y}_{i}\right)^{2}
\end{equation}

With the S-G smoothing filter, on the other hand, the estimates of
the signal, $\widehat{y}_{i}$, are obtained from local models with
little or no a priori knowledge to inform the selection of the parameters
of the local model. For a given number of polynomial coefficients,
\emph{n}, a sample size that is too small will result in over-fitting
and a noise variance estimate from Equation 3.2 that is too small
while a sample size that is too large will result in under-fitting
and a noise variance estimate that is too large. To address this problem
the following equation for estimating the input noise variance is
proposed:

\begin{equation}
\sigma_{e}^{2}=\frac{1}{2\,(q-1)}\,\stackrel[i=1]{q-1}{\sum}\left[(y_{i+1}-y_{i})-(\widehat{y}_{i+1}-\widehat{y}_{i})\right]^{2}
\end{equation}

The filter input, $y_{i}$, has two additive components, signal and
noise. The differencing of the filter input, $(y_{i+1}-y_{i})$, gives
the noise plus the trend in the signal. The trend in the signal is
estimated by differencing the filter output, $(\widehat{y}_{i+1}-\widehat{y}_{i})$.
If the estimate of the trend in the signal is reliable then subtracting
the trend in the signal from the noise plus the trend in the signal
leaves mostly noise with very little signal. The expression in square
brackets can be rearranged as $(y_{i+1}-\widehat{y}_{i+1})-(y_{i}-\widehat{y}_{i})$.
For uncorrelated noise the variance of this quantity is twice the
noise variance hence the compensating factor of two in Equation 3.3.

Equations 3.2 and 3.3 give biased estimates of the input noise variance.
For the purpose of calculating confidence intervals it is better to
use unbiased estimates. The following equation, derived in Appendix
A, corrects for the degrees of freedom, $2m+1-n$, associated with
the polynomial fits to give unbiased estimates:

\begin{equation}
\sigma_{\mathrm{unbiased}}^{2}=\frac{2m+1}{2m+1-n}\:\sigma_{\mathrm{biased}}^{2}
\end{equation}

Because the filter is linear and if the filter parameters are selected
to minimize distortion of the signal, the filter's action on the noise
can be analysed independently of the signal. The noise component of
the filter output, $\widehat{y}$, is given by $\widehat{e}$:

\begin{equation}
\widehat{e}=c_{1}\,e_{1}+c_{2}\,e_{2}+\cdots+c_{2m+1}\,e_{2m+1}
\end{equation}

If the filter input noise, \emph{e}, is of constant variance, $\sigma_{e}^{2}$,
then the variance of the filter output noise, $\sigma_{\widehat{e}}^{2}$,
is given by:

\begin{equation}
\sigma_{\widehat{e}}^{2}=\sigma_{e}^{2}\stackrel[i=1]{2m+1}{\sum}c_{i}^{2}
\end{equation}

If the input noise, \emph{e}, is normally distributed then the 95\%
confidence intervals for the values of the filter output are $\left[\widehat{y}_{i}-1.96\,\sigma_{\widehat{e}},\,\widehat{y}_{i}+1.96\,\sigma_{\widehat{e}}\right]$.
Note that Equation 3.6 also applies more generally for any statistic
that is a linear combination of the filter inputs. Therefore it applies
to the derivatives of the filter output (Equation 2.15) giving estimates
of the variance of the derivatives and the associated confidence intervals.\newpage{}

An analysis of the properties of the S-G filter shows that the number
of polynomial parameters, \emph{n}, should be an odd number. If \emph{n}
is an odd number then using $n+1$ will give the same values of $\widehat{y}_{i}$
between the tails of the data set but give larger confidence intervals
for the values of $\widehat{y}_{i}$ in the tails. Furthermore using
$n+1$ will give larger confidence intervals for the derivatives,
$d\widehat{y}_{i}/dx$ for the entire data set and significantly larger
confidence intervals in the tails.

Applying Equation 3.3 is complicated by the fact that the trend in
the signal, $(\widehat{y}_{i+1}-\widehat{y}_{i})$, depends on the
filter parameters and the values of these parameters that give a reliable
estimate of the trend in the signal are a priori unknown. The resolution
of this complication will be demonstrated in the next section with
a statistical analysis of the Keeling data for atmospheric CO\textsubscript{2}
concentrations measured over the last 67 years.

\section{An Analysis of the Keeling Data for Atmospheric CO\protect\textsubscript{2}
Levels}

Charles Keeling, a scientist at the Scripps Institution of Oceanography,
started taking continuous measurements of atmospheric CO\textsubscript{2}
concentration at Hawaii's Mauna Loa Observatory in 1958. This effort
has been continued by others to the present day giving a continuous
record of atmospheric CO\textsubscript{2} concentration spanning
67 years {[}4{]}. Figure 1a, the so-called Keeling curve, shows the
annual average measured CO\textsubscript{2} concentration in parts
per million (ppm), $y_{i}$, and Figure 1b shows the annual change
in the average measured CO\textsubscript{2} concentration in ppm,
$y_{i+1}-y_{i}$ or $\Delta y$. This rate of change in CO\textsubscript{2}
concentration is of particular interest to climatologists. The trend
in the rate of change is roughly linear as shown by the red line but
finer features in the trend are somewhat obscured by noise. 
\begin{center}
\includegraphics[scale=0.8]{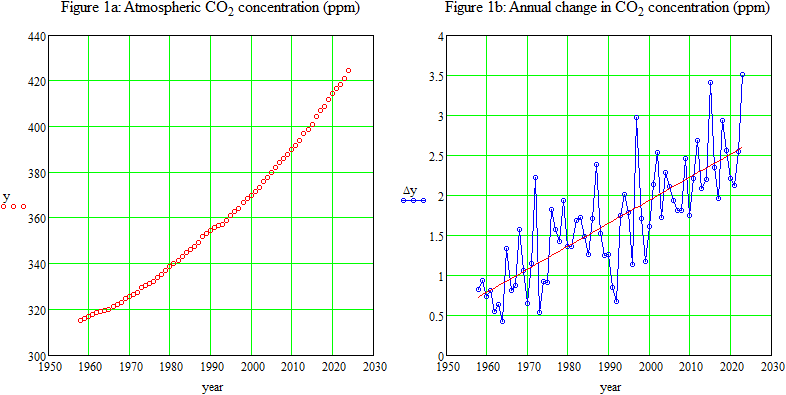}
\par\end{center}

Because the Savitzky-Golay filter can preserve higher moments in the
data it is well suited for filtering noise without filtering finer
features of the signal. The S-G filter was applied to the data of
Figure 1a for values of the filter parameter \emph{n} of 3, 5 and
7 and values of \emph{m} ranging up to 25. The standard deviations
of the S-G filter residuals using Equation 3.2 are plotted in Figure
2a. This figure shows that the estimate of the noise variance given
by the variance of the filter residuals using Equation 3.2 can be
quite sensitive to the values of these parameters. And without a reliable
estimate of the noise variance there is no basis for determining confidence
intervals on the filter output. A clue to solving this problem is
given in Figure 1b showing the annual change in the average measured
CO\textsubscript{2} concentration, $\Delta y$. The effect of differencing
the measurements is to amplify the noise to signal ratio making it
easier to estimate the noise variance. The linear trend in the noise
can be taken to be signal so the noise to signal ratio is further
amplified by subtracting the linear trend. This leaves mostly noise
and very little signal. The standard deviation of this noise after
most of the signal has been removed is 0.32 ppm.\newpage{}

The general idea behind this procedure is formalized in Equation 3.3.
The term $(y_{i+1}-y_{i})$ is the difference in the measurements
and the term $(\widehat{y}_{i+1}-\widehat{y}_{i})$ is the corresponding
estimate of the difference in the signal. When the estimated difference
in the signal is subtracted from the difference in the measurements
the result is an estimate of the noise.
\begin{center}
\includegraphics[scale=0.8]{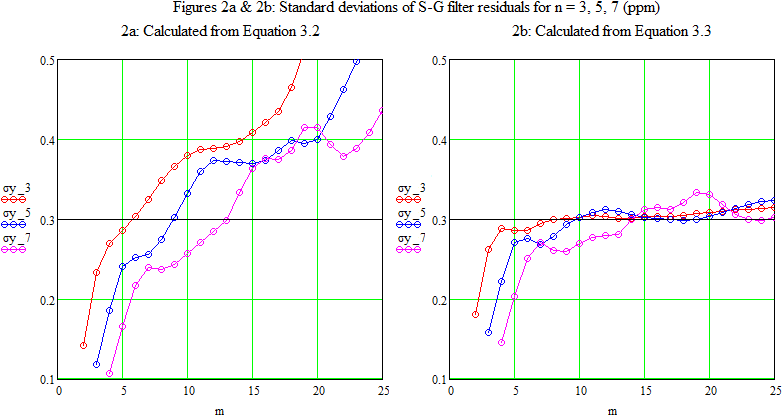}
\par\end{center}

For the same values of the S-G filter parameters as in Figure 2a the
standard deviation of the S-G filter residuals using Equation 3.3
are plotted in Figure 2b. The contrast between the results in these
two figures is significant. The rationale for using Equation 3.3 is
based on the observation that the estimate of the trend in the signal,
the equation term $(\widehat{y}_{i+1}-\widehat{y}_{i})$, is relatively
insensitive to moderate amounts of under-fitting associated with larger
values of the S-G filter parameter \emph{m}. This rationale is clearly
supported by the results shown in Figure 2b where a reliable (but
biased) estimate of the standard deviation of the noise in the Keeling
data is indicated by the horizontal black line at 0.30 ppm. The dropoffs
in the standard deviations for small values of the parameter \emph{m}
are an indication that the data is being over-fitted and the premise
supporting Equation 3.3 does not hold.

For a given value of \emph{n} the appropriate value of \emph{m} would
ideally yield filter residuals with a standard deviation equal to
the estimate of the standard deviation of the noise. Given the discrete
nature of the filter parameters, a reasonable choice for \emph{m}
would correspond to the value of the filter residual standard deviation
that is closest to the estimated noise standard deviation. So for
the values of \emph{n} of 3, 5 and 7 the respective choices of \emph{m}
would be 6, 9 and 13. The use of biased estimates of the standard
deviations in Figures 2a and 2b make it easier to identify the appropriate
value of \emph{m}. And because the corresponding values in the plots
of Figures 2a and 2b are biased by the same factor no error in the
determination of the value of \emph{m} is introduced by the use of
biased estimates. 

The effectiveness of the S-G filter to separate the signal from the
noise increases with the value of \emph{n}. So a value of \emph{n}
as low as 3 might distort the signal because a quadratic polynomial
is not good at fitting features like the inflection points seen in
Figure 5 near the year 1988. However there is also a problem with
choosing a large value of \emph{n}. The uncertainty in the estimate
of the derivative of the signal in the tails of a data set increases
rapidly with increasing values of \emph{n}. As a practical matter,
a value of \emph{n} of 5 (or maybe 7 if there are sharp peaks in the
data) is a good compromise.

With S-G filter parameters of 5 for \emph{n} and 9 for \emph{m} the
biased standard deviation of the filter residuals is 0.301 ppm, very
close to the biased estimate of the standard deviation of the noise,
0.300 ppm from Figure 2b. The unbiased estimate of the residual standard
deviation from Equation 3.4 is 0.351 ppm.\newpage{}

The CO\textsubscript{2} data with the filtered values is given in
Figure 3a and the normalized filter residuals are plotted in Figure
3b. The variable name 'yf' (for y filtered) in the figures is $\widehat{y}$
in the text.
\begin{center}
\includegraphics[scale=0.8]{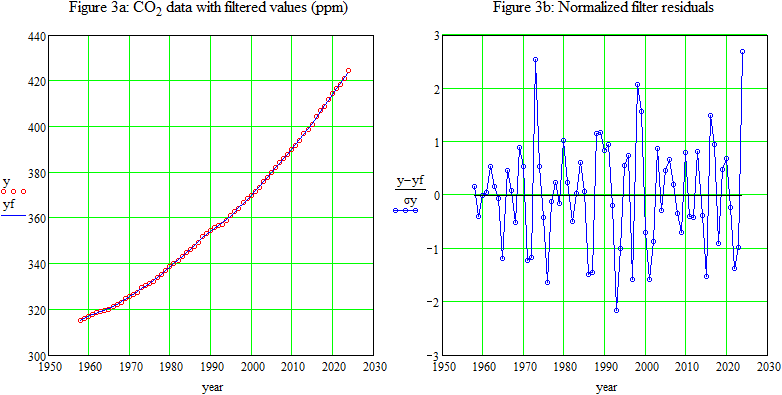}
\par\end{center}

For the purpose of estimating confidence intervals it is assumed that
the noise is normally distributed with zero mean and constant variance.
The assumption of constant variance is checked by comparing the residual
variance for the first half of the data and the second half of the
data. With two sample sizes of 34 the variance ratio of 0.64 is within
the 95\% confidence interval for the F test, supporting the assumption
of constant variance.

A graphical test for normality is given by the normal probability
plot of Figure 4a. This plot doesn't indicate a significant departure
from a normal distribution. Polynomials with degrees ranging from
2 to 20 were fit to the 67 values of the Keeling data. The unbiased
standard deviations of the residuals for these fits are plotted in
Figure 4b where the minimum value is 0.358 ppm. This close agreement
with the unbiased estimate of 0.351 ppm calculated using the S-G filter
with Equations 3.3 and 3.4 and indicated by the black line lends support
to the methods proposed in this study.
\begin{center}
\includegraphics[scale=0.8]{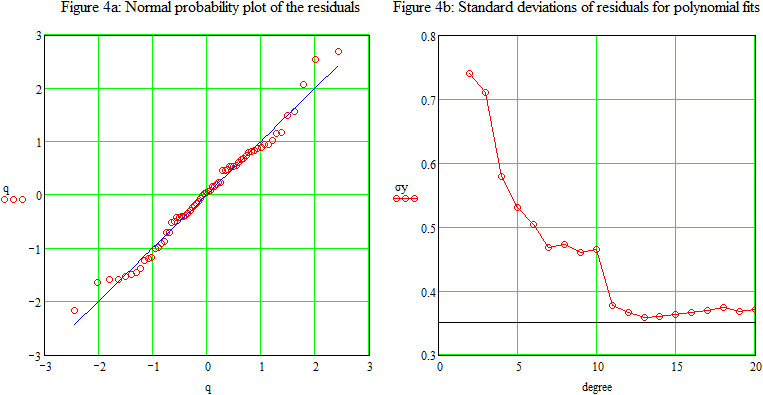}
\par\end{center}

Climatologists are particularly interested in the rate of change of
the accumulation of CO\textsubscript{2} in the atmosphere. The estimated
derivative of a signal is given by Equations 2.14 and 2.15 and confidence
intervals on the estimated derivative are based on Equation 3.6. The
plot of the estimated derivative with 95\% confidence intervals is
shown in Figure 5 where the variable dyf (i.e., $d\widehat{y}/dx$)
denotes the filtered annual change in ppm/year. The widely scattered
points, $\Delta\mathrm{y}$, are the estimates of annual change from
simple differencing of the annual values as plotted in Figure 1b.
\begin{center}
\includegraphics[scale=0.8]{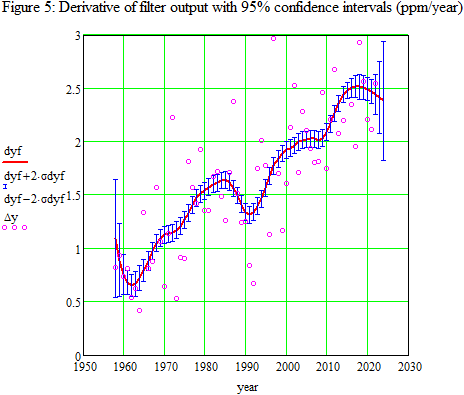}
\par\end{center}

A Monte Carlo simulation was conducted to verify that the confidence
intervals are realistic. The filter input, $y$, was taken as the
signal. Noise was added to this signal where the noise was randomly
sampled from a normal distribution with a standard deviation of 0.351
ppm. The S-G filter with filter parameters $n=5$ and $m=9$ was then
applied to the noisy data. The differences between the derivatives
of the signal and the derivatives of the filtered output (Equation
2.15) for the 67 data points were then calculated. This was repeated
1000 times and the standard deviations of the calculated differences
are plotted in Figure 6 in red along with the estimated standard deviations
from Equation 3.6 in blue. The close agreement suggests that confidence
intervals derived from Equation 3.6 are realistic.
\begin{center}
\includegraphics[scale=0.8]{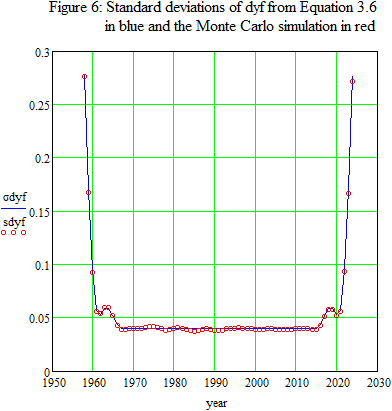}
\par\end{center}

\section{A Further Analysis of the Keeling Data for Atmospheric CO\protect\textsubscript{2}
Levels}

Climatologists have determined that the preindustrial value of atmospheric
CO\textsubscript{2} concentration was about 280 ppm and that the
value has not exceeded 300 ppm in the last 800,000 years {[}6{]}.
So values in excess of 280 ppm can be attributed to human activity,
especially the combustion of fossil fuels. Figure 7 shows the logarithm
base 2 of the CO\textsubscript{2} concentration in excess of 280
ppm for the Keeling data (red) and the S-G filtered data (blue). The
trend is roughly linear and the average doubling period is 33 years.
\begin{center}
\includegraphics[scale=0.8]{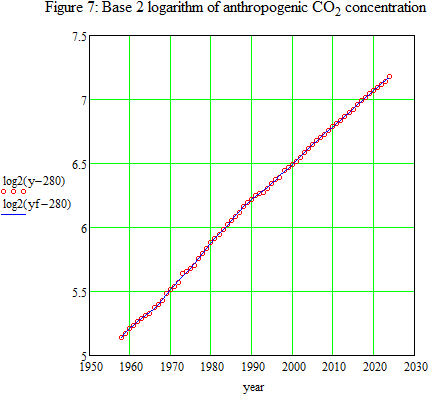}
\par\end{center}

Figure 8 shows the annual fractional rate of change in the anthropogenic
component of the CO\textsubscript{2} concentration. The average fractional
rate of change over the range of the data is 0.021 per year or 2.1\%
per cent per year.
\begin{center}
\includegraphics[clip,scale=0.8]{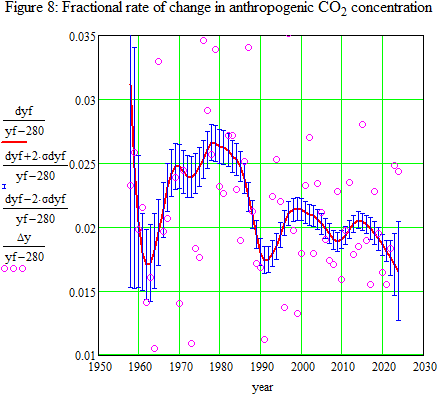}
\par\end{center}

\noindent The fractional rate of change in anthropogenic CO\textsubscript{2}
concentration peaked at about 2.6\% per year around 1980 with a statistically
significant overall downward trend since then. As indicated by the
confidence intervals there is a good deal of uncertainty at the beginning
and end of the data. One factor that is influencing the overall downward
trend in Figure 8 is illustrated in Figure 9. The red curve is the
filtered fractional rate of change in anthropogenic CO\textsubscript{2}
concentration as in Figure 8. The blue curve is the filtered fractional
rate of change in world population over the same time period {[}7{]}.
\begin{center}
\includegraphics[clip,scale=0.8]{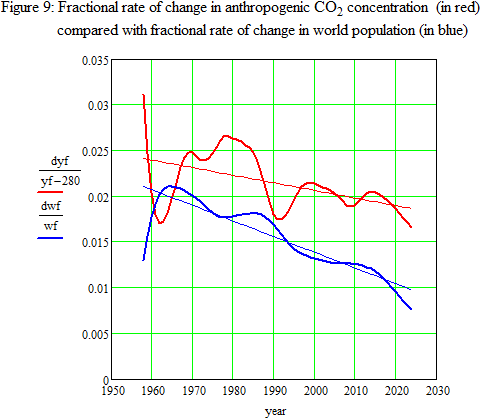}
\par\end{center}

\noindent The thin lines are representations of the overall trends
in the fractional rates of change since 1958. Two trends are influencing
the trend in the fractional rate of change in anthropogenic CO\textsubscript{2}
concentration. The predominant trend, shown in Figure 9, is the decrease
in the fractional rate of change in world population. If the linear
trend is extrapolated it reaches a value of zero world population
growth in 2080. The other trend that has offset the influence of the
trend in world population has been an increase in the per capita consumption
of fossil fuels in countries with rapid industrial growth over the
last few decades.

\section{Conclusion}

The purpose of this paper has been to challenge the view that ``the
smoothing of data lies in a murky area, beyond the fringe of some
better posed, and therefore more highly recommended {[}statistical{]}
techniques''. Towards this end it has been shown how the Savitzky-Golay
filter can be used to calculate a useful estimate of the noise variance
in a data set. It is then shown how this estimate can be used to determine
filter parameters that generate a filter output that neither under-fits
nor over-fits the data. This, in turn, permits the calculation of
reliable confidence intervals on the filter output and its derivative.
It is this last feature in particular that puts the S-G filter on
a statistically sound foundation.\newpage{}

\section*{References}

{[}1{]} Savitzky, A. and Golay M.J.E. ``Smoothing and differentiation
of data by simplified least-squares

\qquad{}procedures'' \emph{Anal. Chem.} 36 (1964) 1627-39

{[}2{]} Press, W.H et. al. ``Numerical Recipes, 2nd Ed.'' Cambridge
University Press, 2007, page 644

{[}3{]} Orfanidis, S.J. ``Applied Optimal Signal Processing'' McGraw-Hill,
2018, Chapter 3

{[}4{]} National Oceanic and Atmospheric Administration \href{https://gml.noaa.gov/ccgg/trends/data.html}{Link}

{[}5{]} Oxby, P.W. ``An Optimal Weighting Function for the Savitzky-Golay
Filter'' (2021) \href{https://arxiv.org/abs/2111.11667v1}{Link}

{[}6{]} Lindsey, R. et al. ``Climate Change: Atmospheric Carbon Dioxide''
NOAA publication \href{https://www.climate.gov/news-features/understanding-climate/climate-change-atmospheric-carbon-dioxide}{Link}

{[}7{]} Wikipedia ``World Population'' \href{https://en.wikipedia.org/wiki/World_population}{Link}

\section*{Contact}

The author can be contacted at pwoxby@telus.net

\section*{Appendix A - The Rationale in Support of Equation 3.4}

For a least-squares fit of a polynomial with \emph{n} parameters to
a data set with \emph{p} values of a variable \emph{y} and assuming
that the noise in the data is independent and identically distributed
(i.i.d.), a biased estimate of the variance $\sigma^{2}$ of the noise
in the data is given by:

\[
\sigma_{\mathrm{biased}}^{2}=\frac{1}{p}\stackrel[i=1]{q}{\sum}\left(y_{i}-\widehat{y}_{i}\right)^{2}
\]

where $\widehat{y}_{i}$ is the estimate of the signal in the data
obtained from the polynomial fit. An unbiased estimate of the variance
of the noise in the data is given by:

\[
\sigma_{\mathrm{unbiased}}^{2}=\frac{1}{p-n}\stackrel[i=1]{q}{\sum}\left(y_{i}-\widehat{y}_{i}\right)^{2}
\]

This can be rewritten as:

\[
\sigma_{\mathrm{unbiased}}^{2}=\frac{p}{p-n}\:\sigma_{\mathrm{biased}}^{2}
\]

For the S-G filter polynomials with \emph{n} parameters are fitted
to data segments with $2m+1$ values of the variable\emph{ y}. Substituting
$2m+1$ for \emph{p} gives Equation 3.4:

\[
\sigma_{\mathrm{unbiased}}^{2}=\frac{2m+1}{2m+1-n}\:\sigma_{\mathrm{biased}}^{2}
\]

\newpage{}

\section*{Appendix B - Computer Programs}

The programs in this appendix are written in Mathcad code which, by
design, resembles pseudocode. They take as input the number of parameters,
\emph{n}, in the fitting polynomial, the number of data samples, $2m+1$,
to which the polynomial fit is applied, and the filter input data,
a vector \emph{y} of \emph{q} values evenly spaced on the x-axis.

The following two programs, XMA(n,m), both calculate the matrix $X\left(X^{T}WX\right)^{-1}X^{T}W$
of Equation 2.10 and the matrix $XD\left(X^{T}WX\right)^{-1}X^{T}W$
of Equation 2.14. The only difference between the two programs is
that one program explicitly subscripts all vectors and matrices whereas
the other program uses a compact matrix notation where appropriate.
The 'M' in XMA represents either the identity matrix \emph{I} implicit
in Equation 2.10 or the matrix \emph{D} in Equation 2.14.
\begin{center}
\includegraphics[scale=0.8]{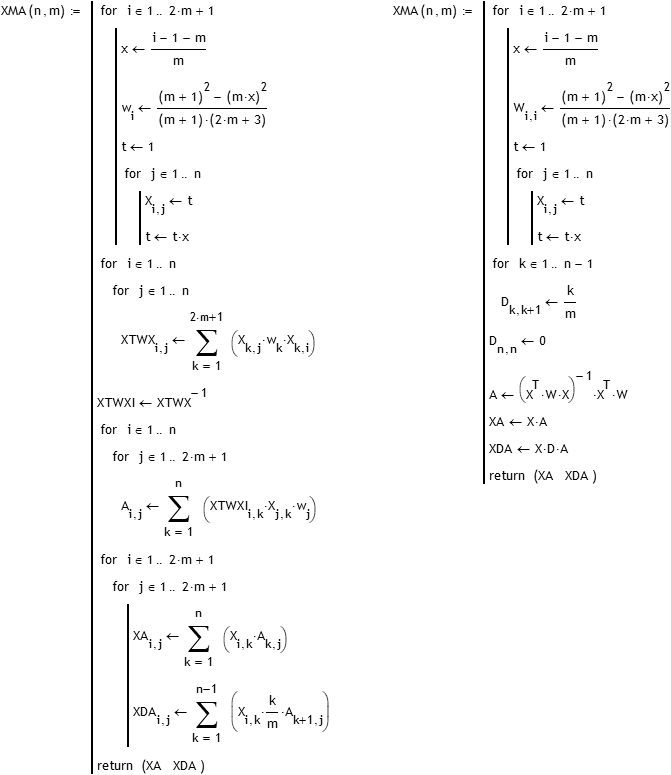}\newpage{}
\par\end{center}

The program SG(n,m,q,y) takes as input the S-G filter parameters,
\emph{n} and \emph{m}, and the number of elements, \emph{q}, in the
vector of data values, \emph{y}, to be filtered. The filter coefficients
are contained in the rows of the matrices \emph{XA} and \emph{XDA}
calculated by the program XMA(n,m). Equation 2.11 is used to calculate
the vector \emph{yf} (i.e., $\widehat{y}$) of filtered values and
Equation 2.15 is used to calculate the vector \emph{dyf} (i.e., $d\widehat{y}/dx$)
of the first derivatives of the filtered values. Then Equation 3.6
is used to calculate the standard deviations, \emph{syf} , of the
filtered values and the standard deviations, \emph{sdyf} , of the
derivatives of the filtered values. The confidence intervals are based
on the standard deviations and are approximately plus or minus twice
the standard deviations for 95\% confidence intervals.

The program SD(n,p,q,y) calculates the standard deviations, \emph{sd},
of the filter residuals used to create the plots as illustrated in
Figures 2a and 2b. The program takes as input the filter parameter\emph{
n} and the maximum value, \emph{p}, of the filter parameter \emph{m}
that is to be plotted on the x-axis of the figures. The minimum value
of \emph{m} is constrained by the inequality $2m+1>\mathrm{n}$. The
last two program inputs are the number of elements, \emph{q}, in the
vector of data values, \emph{y}, to be filtered. The program outputs
the vectors of values of \emph{m}, the residual standard deviations
, \emph{sda}, using Equation 3.2 and the residual standard deviations,
\emph{sdb}, using Equation 3.3. The plot of \emph{sdb} vs \emph{m}
(Figure 2b) is used to estimate the standard deviation of the noise
in the filter input, \emph{y}. The plot of \emph{sda} vs \emph{m}
(Figure 2a) is then used to determine the value of \emph{m} whose
corresponding value of the residual standard deviation,\emph{ sda},
comes closest to the estimated standard deviation of the noise. Note
that \emph{n} is a scalar so SD(n,p,q,y) has to be called for each
of the chosen values of \emph{n} which are 3, 5 and 7 in Figures 2a
and 2b. Note also that the correction for degrees of freedom, Equation
3.4, is not used to calculate the residual standard deviations. The
use of biased estimates makes it easier to identify the appropriate
value of \emph{m}. And because the corresponding values in the plots
of \emph{sda} and \emph{sdb} are biased by the same factor no error
in the determination of the value of \emph{m} is introduced by the
use of biased estimates. 
\begin{center}
\includegraphics[scale=0.8]{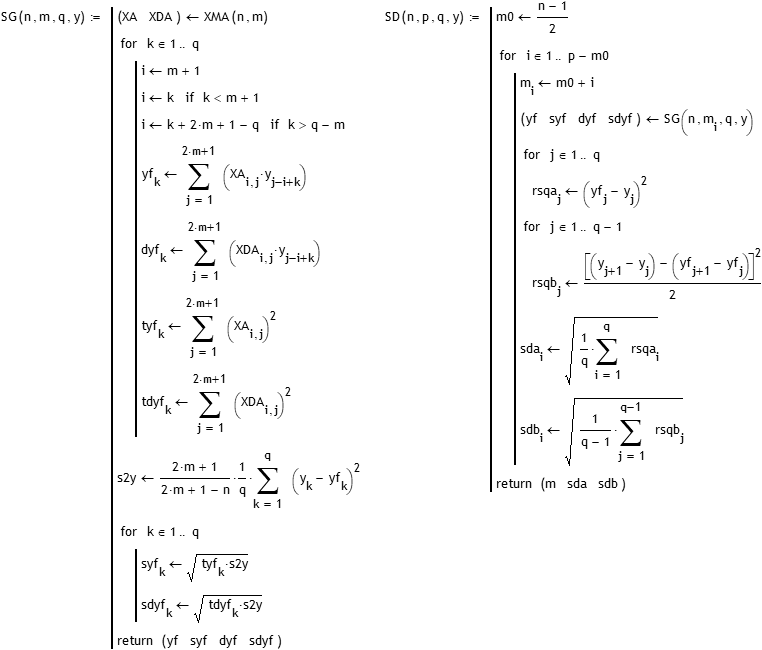}
\par\end{center}
\end{document}